\documentclass{optica-article}

\articletype{Research Article}

\makeatletter
\providecommand{\@journalname}{}
\makeatother
\usepackage{amsmath}

\usepackage{graphicx}
\usepackage{lineno}

\usepackage{lineno}

\begin{document}

%\title{The Role of Low-order Corrections for the Electric Field of Twisted Light\\ in the Paraxial Regime}
%\title{A Dual-Symmetrized Iterative Scheme for Describing Electromagnetic Beams Beyond the Paraxial Approximation}

\title{Dual-Symmetrized Construction of Electromagnetic Beams Beyond the Paraxial Approximation with an Explicit Separation of Scalar and Vectorial Corrections}

\author{Abdullah F. Alharbi,\authormark{1,*} and Kayn A. Forbes,\authormark{2}}

\address{\authormark{1}King Abdulaziz City for Science and Technology (KACST), P.O. Box 6086, Riyadh 11442, Saudi Arabia
\authormark{2}School of Chemistry, University of East Anglia,
Norwich Research Park, Norwich NR4 7TJ, United Kingdom}

\email{\authormark{*}harbi@kacst.edu.sa} %% email address is required; see note below about the corresponding author designation
% use {asbstract*} to suppress the copyright line. Copyright information will be added in production

\section*{Abstract}

We extend a direct Maxwell-iteration procedure for electromagnetic beams beyond the paraxial approximation by consistently retaining the slow longitudinal derivative of the paraxial envelope, incorporating electric–magnetic symmetrization and a scalar-envelope correction. The present formulation provides a direct field-level derivation of the dual-symmetrized vectorial corrections through successive applications of Maxwell’s equations. The complementary electric-first and magnetic-first field pairs are subsequently averaged, thereby removing the construction-dependent assignment of the vectorial correction to either field. In addition, the scalar envelope correction is obtained independently from the isotropic co-polar contribution to the solutions of the higher-order Helmholtz equation. The resulting corrected fields apply to an arbitrary paraxial envelope and input Jones polarization.
\section{Introduction}

The paraxial approximation remains the standard framework for describing weakly focused optical beams. In this approximation, the field is treated as transverse to the propagation direction, and the spatial envelope is assumed to vary slowly along the optical axis. This description is extremely successful for broad beams with waists much larger than the wavelength. However, it becomes incomplete when the beam is tightly confined. In such regimes, the electromagnetic field can no longer be treated as purely transverse: longitudinal components appear, cross-polarized transverse components are induced, and spin-orbit effects become important \cite{Bliokh2010SpinOrbit,Schmiegelow_Transfer_2016,Quinteiro_twisted_2017,Conry_CrossPolarization_2012}.

A useful starting point for describing fields beyond the paraxial approximation is the observation, emphasized by Lax, Louisell, and McKnight, that the usual paraxial beam ansatz is not compatible with Maxwell's equations \cite{Lax_From_1975}. 
Perturbative schemes address this issue by treating the paraxial field as the zeroth-order term in an expansion in the small parameter
\begin{align}
f=\frac{1}{kw_0},
\end{align}
where $k$ is the wavenumber and $w_0$ is the beam waist. Longitudinal corrections occur at odd orders in $f$, while transverse corrections occur at even orders \cite{Lax_From_1975}. Thus, the leading transverse paraxial field is denoted $T0$, the first longitudinal correction is denoted $L1$, and the second transverse correction is denoted $T2$.

Several perturbative constructions have appeared in the literature, differing in their starting assumptions and boundary conditions \cite{Davis_Theory_1979,Agrawal_Gaussian_1979,Agrawal_Free_1983,KIM_Hermite_1999,Kim_HigherOrder_1999,Chen_analyses_2002,Salamin_Gaussian_2007,Vovk_Analytical_2017}. Two principal approaches can be identified in the early literature, distinguished by the type of electromagnetic beam they generate: electric-type and magnetic-type beams \cite{Sheppard_Electromagnetic_1999}. In Davis' formulation, the magnetic vector potential is chosen to be plane-polarized, leading to a comparatively simple magnetic field while placing the higher-order vectorial structure primarily in the electric field \cite{Davis_Theory_1979}. In contrast, the Agrawal-Lax construction fixes the transverse electric field more rigidly, effectively shifting the higher-order vectorial structure into the magnetic field \cite{Agrawal_Free_1983}. These constructions agree at first order, but differ at second order and beyond.
In particular, the electric-type formulation allows higher-order transverse field components to possess transverse polarizations that differ from that of the zeroth-order field. This feature is consistent with full numerical solutions, which result in a rich and spatially varying polarization structure in tightly focused beams. By contrast, the magnetic-type approach constrains the transverse polarization more rigidly, suppressing some of these effects at the same order.

Despite this qualitative advantage,  Davis' formulation lacks symmetry between the electric and magnetic field components beyond first order. To restore this symmetry, Barton and Alexander introduced a dual vector-potential construction, generating a second electromagnetic solution complementary to the original one \cite{Barton_fifth_1989}.
As a result, the two solutions are averaged to recover a balanced electromagnetic beam.

An interesting iterative procedure has recently been used to generate higher-order corrections to paraxial vortex beams directly from Maxwell’s equations. In this method, one begins with a transverse paraxial electric field, imposes Gauss's law to obtain the first longitudinal electric component, and then applies the curl equations to generate the next transverse corrections \cite{Forbes_VortexLightNanoscale_2025}. This method can be viewed as a variant of Lax’s original formulation \cite{Lax_From_1975}. Despite its physical transparency, the reported procedure neglects the slow longitudinal derivative of the paraxial envelope, even though this derivative contributes at second order in the paraxial parameter. When this contribution is retained, the second-order vectorial correction, for example, is assigned entirely to the electric field. An opposite artificial imbalance arises when a magnetic-first construction is chosen, in which the second-order vectorial correction is instead assigned entirely to the magnetic field. 

%Another significant limitation of the reported iterative method is that it keeps the paraxial envelope fixed throughout the calculation, without incorporating higher-order scalar-envelope corrections.

Moreover, such a Maxwell-iteration procedure takes the paraxial envelope as fixed throughout the construction. 
Neglecting the scalar nonparaxial correction to the slowly varying envelope leaves the co-polar transverse amplitude and phase profile incomplete. As a result, quantities that depend on the detailed field structure, such as intensity distributions, polarization properties, and light–matter interaction strengths can be inaccurately calculated.

%\textcolor{red}{Barton and Alexander [15] introduced a dual-vector-potential construction in which complementary electric and magnetic solutions are averaged to obtain dual-symmetrized fields. The present work adopts this established symmetrization principle but constructs the leading vectorial corrections directly at the field level through successive application of Maxwell’s divergence and curl equations. More recently, an iterative procedure was used to derive higher-order corrections to paraxial vortex beams directly from Maxwell’s equations [16]. Although physically transparent, the simplified procedure neglects the slow longitudinal derivative of the paraxial envelope, which contributes at order \(f^{2}\). Furthermore, an iteration performed with a fixed paraxial envelope does not determine the additional scalar-envelope correction. These limitations concern the implementation of the simplified iteration, rather than the availability of scalar corrections or symmetric field constructions in nonparaxial beam theory.
Here, we extend this iterative procedure by retaining the slow longitudinal derivative, combining complementary electric-first and magnetic-first constructions, and incorporating the scalar-envelope correction under a specified boundary or asymptotic prescription. The contribution lies in bringing these ingredients together within an order-consistent field-level formulation and identifying the polarization-isotropic and vectorial contributions separately. This explicit separation clarifies the origin and construction dependence of the correction terms; it does not introduce symmetrization or the scalar–vectorial distinction as new principles. We illustrate the formulation for Gaussian and ($p=0$) Laguerre–Gaussian beams and establish its connection to earlier results.

In this work, we develop an iterative scheme that incorporates both ingredients consistently: the vectorial corrections imposed by Maxwell’s equations and the scalar nonparaxial correction to the envelope.
The scheme yields a more accurate and balanced field description of the former by  retaining the slow longitudinal derivative and adopting the symmetrization principle. However, in contrast to the dual-vector-potential construction of Barton and Alexander, the present work constructs the dual-symmetrized vectorial corrections by working directly with the electric and magnetic fields and applying Maxwell’s divergence and curl equations successively. We also clarify that a field iteration performed with a fixed paraxial envelope does not, by itself, determine the additional scalar-envelope correction. We obtain this correction from the expanded Helmholtz equation and identify its polarization-isotropic contribution separately from the vectorial terms. We illustrate the formulation for Gaussian and \(p=0\) Laguerre–Gaussian (LG) beams and establish its connection to earlier results.

\section{Theoretical Background}

We consider a monochromatic beam propagating in a homogeneous, isotropic, lossless medium. 
Assuming harmonic time dependence $\exp(-i\omega t)$, Maxwell's equations become

\begin{align}
\nabla\cdot\mathbf E &=0, 
\label{eq:gaussE}
\\
\nabla\cdot\mathbf B &=0,
\label{eq:gaussB}
\\
\nabla\times\mathbf E &= ikc\,\mathbf B,
\label{eq:faraday}
\\
\nabla\times\mathbf B &
=-\frac{ik}{c}\mathbf E.
\label{eq:ampere}
\end{align}

We introduce the dimensionless variables
\begin{align}
\xi=\frac{x}{w_0},
\qquad
\eta=\frac{y}{w_0},
\qquad
\zeta=\frac{z}{kw_0^2},
\qquad
f=\frac{1}{kw_0}.
\end{align}
The carrier factor is
\begin{align}
e^{ikz}=e^{i\zeta/f^2}.
\end{align}
The derivatives transform as
\begin{align}
\frac{\partial}{\partial x}
&=
kf\frac{\partial}{\partial \xi},
\\
\frac{\partial}{\partial y}
&=kf\frac{\partial}{\partial \eta},
\\
\frac{\partial}{\partial z}
\left[e^{i\zeta/f^2}\psi
\right]
&=k e^{i\zeta/f^2}\left(i\psi+f^2\frac{\partial\psi}{\partial \zeta}
\right).
\label{eq:z_derivative_rule}
\end{align}
Equation~\eqref{eq:z_derivative_rule} is important: transverse derivatives bring one power of $f$, while the slow longitudinal derivative of the envelope enters at order $f^2$.

%======================

\begin{figure}[t]
    \centering
    \includegraphics[width=0.6\linewidth]{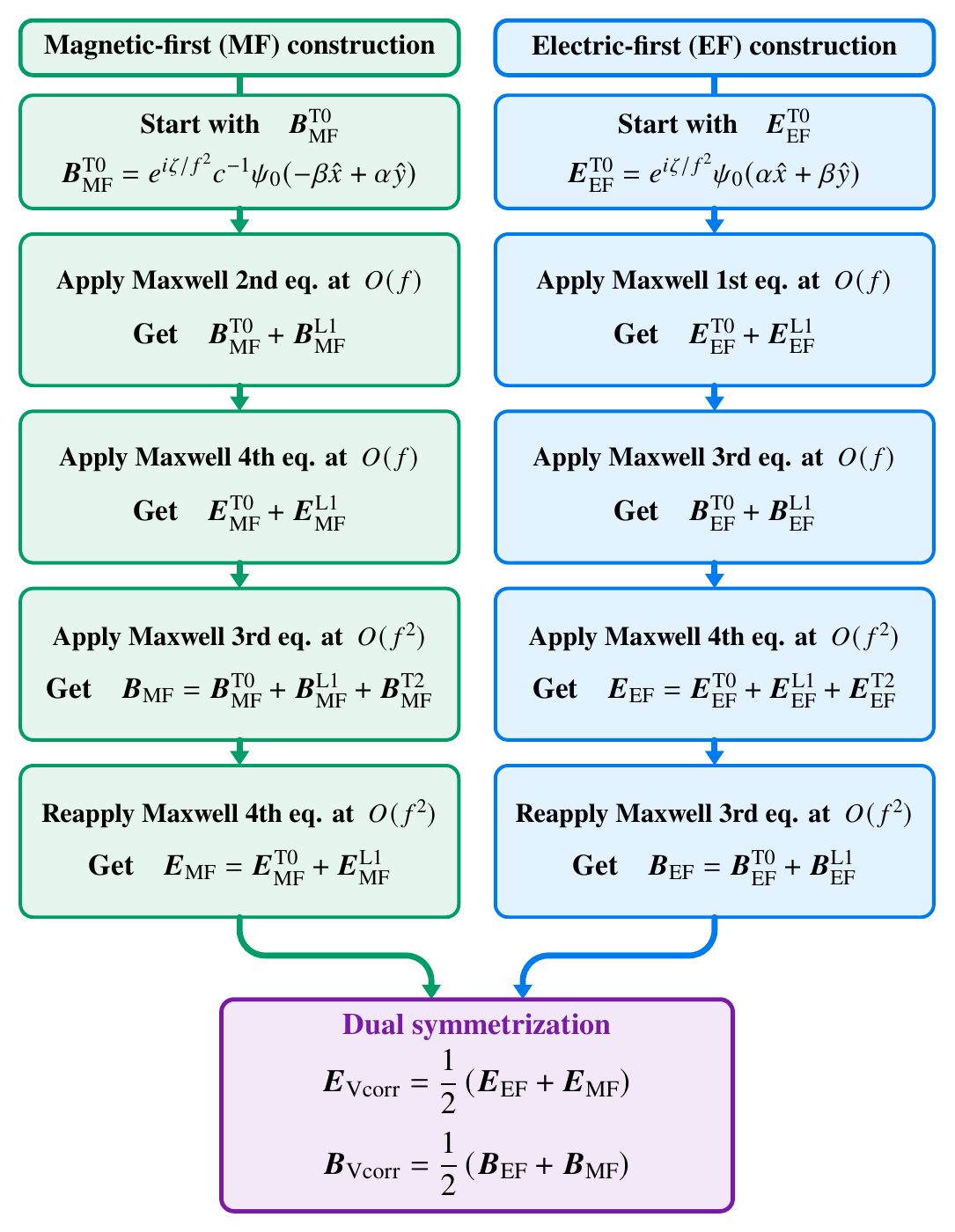}
    \caption{Flowchart for deriving
    $\mathbf{E}_{\mathrm{Vcorr}}$ and
    $\mathbf{B}_{\mathrm{Vcorr}}$
    through second order in $f$ by averaging the
    magnetic-first (MF) and electric-first (EF)
    Maxwell constructions.}
    \label{fig:Flowchart}
\end{figure}

\section{Vectorial Correction and Symmetrization}

 To derive the fields through successive applications of Maxwell's equations, we first adopt the electric-first approach. In this construction, the electric field is corrected through order $f^n$ before the corresponding magnetic field is evaluated to the same order. The starting point is the transverse paraxial electric field:

\begin{align}
\mathbf E_{\mathrm{EF}}^{T0}
=
e^{i\zeta/f^2}
\psi_0\left(\alpha\hat{\mathbf x}
+\beta\hat{\mathbf y}\right),
\label{eq:E_T0}
\end{align}

where $\alpha$ and $\beta$ are the Jones-vector coefficients of the input polarization.

Because the transverse field in Eq.~\eqref{eq:E_T0} has nonzero transverse divergence, Gauss's law requires a longitudinal electric component. We write
\begin{align}
\mathbf E_{\mathrm{EF}}^{L1}
=
e^{i\zeta/f^2}
E_{z,\mathrm{EF}}^{L1}
\hat{\mathbf z}.
\end{align}
Using Eq.~\eqref{eq:gaussE} and keeping the leading carrier derivative in $\partial E_z^{L1}/\partial z$, we obtain
\begin{align}
E_{z,\mathrm{EF}}^{L1}=if
\left(\alpha\frac{\partial\psi_0}{\partial \xi}
+\beta\frac{\partial\psi_0}{\partial \eta}
\right).
\label{eq:Ez_L1}
\end{align}
Therefore,
\begin{align}
\mathbf E_{\mathrm{EF}}^{T0+L1}
=
e^{i\zeta/f^2}
\left[\alpha\psi_0\hat{\mathbf x}
+\beta\psi_0\hat{\mathbf y}
+if\left(\alpha\frac{\partial\psi_0}{\partial \xi}
+\beta\frac{\partial\psi_0}{\partial \eta}
\right)
\hat{\mathbf z}
\right].
\label{eq:E_T0_L1}
\end{align}

Next, the magnetic field, to first order, is obtained by applying Faraday's law to Eq.~\eqref{eq:E_T0_L1}:

\begin{align}
\mathbf B_{\mathrm{EF}}^{T0+L1}
=
\frac{e^{i\zeta/f^2}}{c}
\left[
-\beta\psi_0\hat{\mathbf x}
+\alpha\psi_0\hat{\mathbf y}
+if\left(\alpha\frac{\partial\psi_0}{\partial \eta}
-\beta\frac{\partial\psi_0}{\partial \xi}
\right)\hat{\mathbf z}
\right].
\label{eq:B_T0_L1}
\end{align}

We now insert Eq.~\eqref{eq:B_T0_L1} into Maxwell-Ampere's law ~\eqref{eq:ampere} while
keeping the slow longitudinal derivative of $\psi_0$. Thus, the electric-first electric field becomes the following:

\begin{align}
\mathbf E_{\mathrm{EF}}
=e^{i\zeta/f^2}\Bigg[
&
\left\{\alpha\psi_0
+f^2\left[-\alpha\frac{\partial^2\psi_0}{\partial \eta^2}
+\beta\frac{\partial^2\psi_0}{\partial \xi\partial\eta}-i\alpha\frac{\partial\psi_0}{\partial \zeta}
\right]\right\}\hat{\mathbf x}\nonumber\\
+
&
\left\{
\beta\psi_0
+f^2\left[-\beta\frac{\partial^2\psi_0}{\partial \xi^2}
+\alpha\frac{\partial^2\psi_0}{\partial \xi\partial\eta}-i\beta\frac{\partial\psi_0}{\partial \zeta}
\right]\right\}\hat{\mathbf y}\nonumber\\+
&
if\left(\alpha\frac{\partial\psi_0}{\partial \xi}+
\beta\frac{\partial\psi_0}{\partial \eta}
\right)\hat{\mathbf z}\Bigg].
\label{eq:E_EF_final}
\end{align}

In deriving Eq. \ref{eq:B_T0_L1}, we applied Faraday’s law only through first order in $f$, as we follow the electric-first approach here. We first update the electric field to order $f^2$, as in Eq. \ref{eq:E_EF_final}, and then reapply Faraday’s law to calculate the magnetic field to the same order. The resulting field $\mathbf B_{\mathrm{EF}}$ has the same form as Eq. \ref{eq:B_T0_L1} with $\mathbf B^{T2}_{\mathrm{EF}}=0$, so the second-order transverse vectorial correction appears only in $\mathbf E_{\mathrm{EF}}$.

In contrast, one ends with a different conclusion when adopting the magnetic-first construction, which begins instead with the leading paraxial magnetic field
\begin{align}
\mathbf B_{\mathrm{MF}}^{T0}
=\frac{e^{i\zeta/f^2}}{c}\psi_0
\left(-\beta\hat{\mathbf x}
+\alpha\hat{\mathbf y}\right).
\end{align}
Imposing $\nabla\cdot\mathbf B=0$ generates the longitudinal magnetic component as in \eqref{eq:B_T0_L1}.
Repeating the same curl iteration gives the complementary result: the second-order vectorial transverse correction appears only in $\mathbf{B}_{\mathbf{MF}}$.

To remove the construction-dependent preference for either starting point and restore electric–magnetic symmetry, we define the dual-symmetrized fields by equally weighting the electric-first and magnetic-first results. 
This choice is motivated by electromagnetic duality: in source-free space, Maxwell’s equations are symmetric under the exchange of the electric and magnetic fields, up to the appropriate factor of ($c$) and sign \cite{Bliokh_Dual_2013}.
Moreover, because both constructions share the same leading-order paraxial field, choosing weights that sum to unity preserves the original normalization. These two requirements lead naturally to equal weights of ($1/2$). Accordingly, 

\begin{align}
\mathbf E_{\mathrm{Vcorr}}
=\frac{1}{2}\left(\mathbf E_{\mathrm{EF}}
+\mathbf E_{\mathrm{MF}}\right),\qquad\mathbf B_{\mathrm{Vcorr}}
=\frac{1}{2}
\left(\mathbf B_{\mathrm{EF}}+\mathbf B_{\mathrm{MF}}\right).
\end{align}
The successive steps of the electric-first and magnetic-first
constructions, together with their subsequent dual symmetrization,
are summarized in the flowchart in Fig.~\ref{fig:Flowchart}. The resulting dual-symmetrized fields become: 

\begin{align}
\mathbf E_{\mathrm{Vcorr}}
=e^{i\zeta/f^2}\Bigg[
&
\left\{\alpha\psi_0
+\frac{f^2}{2}\left[-\alpha\frac{\partial^2\psi_0}{\partial \eta^2}
+\beta\frac{\partial^2\psi_0}{\partial \xi\partial\eta}-i\alpha\frac{\partial\psi_0}{\partial \zeta}
\right]\right\}\hat{\mathbf x}\nonumber\\
+
&
\left\{
\beta\psi_0
+\frac{f^2}{2}\left[-\beta\frac{\partial^2\psi_0}{\partial \xi^2}
+\alpha\frac{\partial^2\psi_0}{\partial \xi\partial\eta}-i\beta\frac{\partial\psi_0}{\partial \zeta}
\right]\right\}\hat{\mathbf y}\nonumber\\+
&
if\left(\alpha\frac{\partial\psi_0}{\partial \xi}+
\beta\frac{\partial\psi_0}{\partial \eta}
\right)\hat{\mathbf z}\Bigg],
\label{eq:E_corr_final}
\end{align}

and
\begin{align}
\mathbf B_{\mathrm{Vcorr}}
=\frac{e^{i\zeta/f^2}}{c}
\Bigg[
&\left\{
-\beta\psi_0+\frac{f^2}{2}\left[\beta\frac{\partial^2\psi_0}{\partial \eta^2}
+\alpha\frac{\partial^2\psi_0}{\partial \xi\partial\eta}+i\beta\frac{\partial\psi_0}{\partial \zeta}
\right]\right\}\hat{\mathbf x}\nonumber\\
+&\left\{\alpha\psi_0
+\frac{f^2}{2}\left[-\alpha\frac{\partial^2\psi_0}{\partial \xi^2}
-\beta\frac{\partial^2\psi_0}{\partial \xi\partial\eta}-i\alpha\frac{\partial\psi_0}{\partial \zeta}
\right]\right\}\hat{\mathbf y}\nonumber\\+
&if\left(\alpha\frac{\partial\psi_0}{\partial \eta}
-\beta\frac{\partial\psi_0}{\partial \xi}\right)\hat{\mathbf z}\Bigg].
\label{eq:B_corr_final}
\end{align}

%We can see that the vectorial second-order correction is divided equally between the electric and magnetic fields.
Although we have presented the construction explicitly up to second order, the same dual-symmetrized logic can be generalized to higher orders.

%\subsection{Example: \(x\)-polarized Gaussian beam}

%======================================
%=====================================
\section{Scalar Correction}

The direct Maxwell iteration keeps the paraxial envelope fixed and only permits vectorial component corrections. In this section, we consult the wave equation to add the scalar correction to the envelope in order to reach the complete form of the corrected fields.

The vector potential $\mathbf A$ is given by the following expression:
\begin{align}
\mathbf A
=
-\frac{i}{\omega} \,e^{i\zeta/f^2}
\psi \,\left(\alpha\hat{\mathbf x}
+\beta\hat{\mathbf y}\right).
\label{eq:A}
\end{align}

The scalar Helmholtz equation for the slowly varying envelope $\psi$ can be written
in the dimensionless variables as
\begin{align}
\left(\nabla^2_\perp
+ 2i\frac{\partial}{\partial \zeta}
+ f^2 \frac{\partial^2}{\partial \zeta^2}\right)\psi=0 .
\label{WaveEquation}
\end{align}

where $\nabla^2_\perp =\frac{\partial^2}{\partial \xi^2}+\frac{\partial^2}{\partial \eta^2}$. In the paraxial approximation, the last term in
Eq.~\eqref{WaveEquation}, which scales as \(f^2\), is neglected and $(\psi=\psi_0) $. 
When corrections of order \(f^2\) are retained, this paraxial solution
is no longer sufficient and the scalar envelope must be
expanded consistently as
\begin{align}
\psi=\psi_0+f^2\psi_2+O(f^4).
\end{align}

Substituting this expansion into Eq.~\eqref{WaveEquation} and collecting
terms order by order gives, in leading order,
\begin{align}
\left(\nabla^2_\perp +2i\frac{\partial}{\partial \zeta}
\right)\psi_0
=
0 ,
\label{eq:paraxial_equation}
\end{align}
which is the paraxial equation. At order \(f^2\), one obtains
\begin{align}
\left(
\nabla^2_\perp+2i\frac{\partial}{\partial \zeta}\right)\psi_2
=-\frac{\partial^2\psi_0}{\partial \zeta^2}.
\label{eq:psi2_equation}
\end{align}

The electric field can be obtained using the Lorenz gauge ($\mathbf E =i\omega (\mathbf A+ k^{-2}\nabla(\nabla.\mathbf{A}))$. Given that the electric-first Maxwell iteration produces the vectorial correction, here we are only interested in the isotropic change of $\mathbf E$ up to second order in $f$:

\begin{equation}
\mathbf{E}_{\text{Scorr}}
=f^{2}e^{i\zeta/f^{2}}
\left(
\psi_{2}\mathbf{e}_{\perp}
+
\left[ \mathbf M\,\mathbf{e}_{\perp}
\right]_{\text{iso}}\right).
\end{equation}

Where $\mathbf{e}_{\perp}=\alpha \hat{\mathbf x}+\beta \hat{\mathbf y}$ and

\begin{align}
\mathbf{M}
&=
\begin{pmatrix}
\partial_{\xi\xi} \psi_0 & \partial_{\xi\eta} \psi_0\\
\partial_{\eta\xi} \psi_0& \partial_{\eta\eta}\psi_0
\end{pmatrix}.\\
&=\left[ \frac{1}{2}
(\mathbf{M}_{xx}+\mathbf{M}_{yy})
\mathbf{I} \right]
+\left[
\mathbf M
-\frac{1}{2}
(\mathbf{M}_{xx}+\mathbf{M}_{yy})
\mathbf{I}
\right].
\end{align}

The first term represents the isotropic part of $\mathbf M$. We can, therefore, write:
\begin{equation}
\mathbf{E}_{\text{Scorr}}
=
f^{2} e^{i\zeta/f^{2}}
\left(\psi_{2}
+\frac{1}{2}\nabla_{\perp}^2\psi_{0}
\right) \mathbf{e}_{\perp}.
\end{equation}

The magnetic field can be obtained from the relation $\mathbf B= \mathbf \nabla \times \mathbf A$. The isotropic change in $\mathbf B$ is:
\begin{equation}
\mathbf{B}_{\text{Scorr}}
=\frac{f^{2}}{c}e^{i\zeta/f^{2}}
\left(\psi_{2}
+\frac{1}{2}\nabla_{\perp}^2\psi_{0}
\right) \hat{\mathbf z} \times\mathbf{e}_{\perp}.
\label{eq:B_scalar}
\end{equation}

In obtaining this expression, we have used the paraxial relation in Eq. \ref{eq:paraxial_equation}.

% Given these results, we can make the following substitution:

% \begin{equation}
% \psi_0 \to \psi_0+\frac{f^2}{2}
% \nabla_{\perp}^{2}\psi_0+ f^2
% \psi_2.
% \end{equation}

% in the expressions \eqref{eq:E_corr_final} and \eqref{eq:B_corr_final} to obtain the final expressions of the corrected fields up to second order in $f$:

After adding the scalar transverse correction, Gauss's law is imposed once more at order $f^3$, which generates the corresponding third-order longitudinal component.
Given these results, the final expressions of the corrected fields up to third order in $f$ can be written as:
\begin{align}
\mathbf{E}_{\mathrm{corr}}
&=\mathbf E_{\text{Vcorr}}+\mathbf E_{\text{Scorr}}\\
&= \, \, e^{i\zeta/f^2}
\Bigg[
\Bigg\{
\alpha\psi_0
+
f^2
\left[
\alpha\psi_2
+
\frac{3\alpha}{4}
\frac{\partial^2\psi_0}{\partial\xi^2}
+
\frac{\alpha}{4}
\frac{\partial^2\psi_0}{\partial\eta^2}
+
\frac{\beta}{2}
\frac{\partial^2\psi_0}{\partial\xi\partial\eta}
\right]
\Bigg\}
\hat{\mathbf{x}}
\nonumber\\
& \,\,\,\,\,\,\,\,\,\,\,\,\,\,\, +
\Bigg\{
\beta\psi_0
+
f^2
\left[
\beta\psi_2
+
\frac{\alpha}{2}
\frac{\partial^2\psi_0}{\partial\xi\partial\eta}
+\frac{\beta}{4}\frac{\partial^2\psi_0}{\partial\xi^2}
+\frac{3\beta}{4}
\frac{\partial^2\psi_0}{\partial\eta^2}
\right]
\Bigg\}
\hat{\mathbf{y}}
\nonumber\\
& \,\,\,\,\,\,\,\,\,\,\,\,\,\,\, +
\, \Bigg\{ if
\left(
\alpha\frac{\partial\psi_0}{\partial\xi}
+
\beta\frac{\partial\psi_0}{\partial\eta}
\right) +
i f^{3} 
\left(
\alpha \frac{\partial}{\partial \xi}
+
\beta \frac{\partial}{\partial \eta}
\right)
\left(
\psi_{2}
+
\frac{1}{4}\nabla_{\perp}^{2}\psi_{0}
\right)\Bigg\}
\hat{\mathbf{z}}
\Bigg],
\label{eq:E_corr}
\end{align}
and
\begin{align}
\mathbf{B}_{\mathrm{corr}}
&=\mathbf B_{\text{Vcorr}}+\mathbf B_{\text{Scorr}}\\
&=\frac{e^{i\zeta/f^2}}{c}
\Bigg[
\Bigg\{
-\beta\psi_0
+f^2\left[-\beta\psi_2
+\frac{\alpha}{2}\frac{\partial^2\psi_0}{\partial\xi\partial\eta}
-
\frac{3\beta}{4}
\frac{\partial^2\psi_0}{\partial\xi^2}
-
\frac{\beta}{4}
\frac{\partial^2\psi_0}{\partial\eta^2}
\right]
\Bigg\}
\hat{\mathbf{x}}
\nonumber\\
& \,\,\,\,\,\,\,\,\,\,\,\,\,\,\, +
\Bigg\{
\alpha\psi_0
+
f^2\left[
\alpha\psi_2
+
\frac{\alpha}{4}
\frac{\partial^2\psi_0}{\partial\xi^2}
+\frac{3\alpha}{4}\frac{\partial^2\psi_0}{\partial\eta^2}
-\frac{\beta}{2}\frac{\partial^2\psi_0}{\partial\xi\partial\eta}
\right]\Bigg\}\hat{\mathbf{y}}
\nonumber\\
& \,\,\,\,\,\,\,\,\,\,\,\,\,\,\, +
\, \Bigg\{ if\left(\alpha\frac{\partial\psi_0}{\partial\eta}
-\beta\frac{\partial\psi_0}{\partial\xi}\right)
+i f^{3} 
\left(
-\beta \frac{\partial}{\partial \xi}
+
\alpha \frac{\partial}{\partial \eta}
\right)
\left(
\psi_{2}
+
\frac{1}{4}\nabla_{\perp}^{2}\psi_{0}
\right)
\Bigg\} \hat{\mathbf{z}}\Bigg].
\label{eq:B_corr}
\end{align}

As a consistency check, substitution of Eqs.\ref{eq:E_corr} and \ref{eq:B_corr} into Maxwell's equations, together with Eqs. \ref{eq:paraxial_equation} and \ref{eq:psi2_equation}, shows that all terms through the retained perturbative orders cancel:
\begin{align}
\nabla \cdot \mathbf{E}_{\mathrm{corr}}
&= O(f^5), \\
\nabla \cdot \mathbf{B}_{\mathrm{corr}}
&= O(f^5),
\end{align}
while
\begin{align}
\nabla \times \mathbf{E}_{\mathrm{corr}}
- ikc\,\mathbf{B}_{\mathrm{corr}}
&= O(f^4), \\
\nabla \times \mathbf{B}_{\mathrm{corr}}
+ \frac{ik}{c}\mathbf{E}_{\mathrm{corr}}
&= O(f^4).
\end{align}
Thus, the corrected fields satisfy the source-free Maxwell equations to the perturbative order retained in the present construction.

At second order, the contribution of $\psi_2$ is purely polarization-isotropic as it modifies only the co-polar transverse field. At third order, however, $\psi_2$ also contributes through its transverse derivatives to the longitudinal field, leading to a more intricate spatial and polarization-dependent correction.
The presented perturbative hierarchy can be, in principle, extended systematically to obtain the corrected electric and magnetic fields to arbitrary order in $f$.

\section{Examples}
\subsection{$x$-polarized Gaussian Beam}

As a simple example, we consider an \(x\)-polarized Gaussian beam ($\alpha=1,\beta=0$). The leading paraxial envelope is expressed as

\begin{align}
\psi_0(\xi,\eta,\zeta)=-iQ\exp\left(i\rho^2Q\right),
\end{align}

where

\begin{align}
Q=\frac{1}{2\zeta-i},\qquad\rho^2=\xi^2+\eta^2.
\end{align}

Equation \eqref{eq:psi2_equation} does not uniquely specify $\psi_2$, an additional boundary or asymptotic prescription is required.
The first possible boundary-value construction is to define the scalar envelope at the waist ($z=0$) using the paraxial Gaussian description $\psi(\rho,0)=\psi_0(\rho,0)$. In this case $\psi_2(\rho,0)=0$ and $\psi_2(\rho,\zeta)$ becomes a propagation-induced scalar correction. The corresponding solution can be obtained as follows. Using \eqref{WaveEquation}, we can rewrite \eqref{eq:psi2_equation} in the following form:

\begin{align}
\left(
\nabla^2_\perp+2i\frac{\partial}{\partial \zeta}\right)\psi_2
&=\frac{1}{4}\nabla^4_\perp \psi_0\\
&= \frac{1}{4} \frac{1}{2i}\left(
\nabla^2_\perp+2i\frac{\partial}{\partial \zeta}\right) (\zeta \nabla^4_\perp \psi_0).
\end{align}

and therefore
\begin{align}
\psi_2
= -\frac{i\zeta}{8}  \nabla^4_\perp \psi_0
\label{eq:psi2_gaussian1}
\end{align}

For the fundamental Gaussian, this evaluates to
\begin{equation}
\psi_{2}^{(0)}
=\left(4i\zeta Q^{2}
-8\zeta\rho^{2}Q^{3}
-2i\zeta\rho^{4}Q^{4}
\right)\psi_{0}.
\end{equation}

This choice has been adopted by several pioneering studies, for example \cite{Agrawal_Gaussian_1979,Takenaka_Propagation_1985}. 
Instead of this well-defined boundary-value construction, the derivation of Davis' form of $\psi_2$ \cite{Davis_Theory_1979} was motivated by the asymptotic behaviour of a diverging spherical wave.  According to this choice, the second-order correction of the scalar envelope is \cite{Davis_Theory_1979}:

\begin{align}
\psi_2^{(1)}=\left(-2iQ-i\rho^4Q^3\right)\psi_0 .
\label{eq:psi2_gaussian2}
\end{align}

There are also other admissible solutions for $\psi_2$. For example, in Seshadri's construction \cite{Seshadri_fUNDAMENTAL_2008}, the solution contains an additional homogeneous radial contribution that is absent from $\psi_2^{(1)}$. This term arises from imposing the full asymptotic amplitude and phase behavior of the beam.

Setting $\alpha=1$ and $\beta=0$, the corrected electric
field through third order is
\begin{align}
\mathbf{E}_{\mathrm{corr}}
=
e^{i\zeta/f^2}
\Bigg\{&
\left[
\psi_0
+
f^2\left(\psi_2+
\left[2iQ-(3\xi^2+\eta^2)Q^2\right]\psi_0\right)\right]\hat{\mathbf{x}}\nonumber
-2f^2\xi\eta Q^2\psi_0\,\hat{\mathbf{y}}
\nonumber\\
&+\left[-2f\xi Q\psi_0
+
f^3\left(i\frac{\partial\psi_2}{\partial\xi}+\left[-4i\xi Q^2+2\xi\rho^2Q^3\right]\psi_0\right)\right]\hat{\mathbf{z}}\Bigg\}.
\end{align}

and
\begin{align}
\mathbf{B}_{\mathrm{corr}}
=
\frac{e^{i\zeta/f^2}}{c}
\Bigg\{&
-2f^2\xi\eta Q^2\psi_0\,\hat{\mathbf{x}}
+\left[\psi_0+f^2\left(\psi_2+\left[2iQ-(\xi^2+3\eta^2)Q^2\right]\psi_0
\right)\right]\hat{\mathbf{y}}\nonumber\\
&+\left[-2f\eta Q\psi_0
+
f^3\left(i\frac{\partial\psi_2}{\partial\eta}+\left[-4i\eta Q^2+2\eta\rho^2Q^3\right]\psi_0\right)\right]\hat{\mathbf{z}}\Bigg\}.
\end{align}

\begin{figure}[t]
    \centering
    \includegraphics[width=0.8\linewidth]{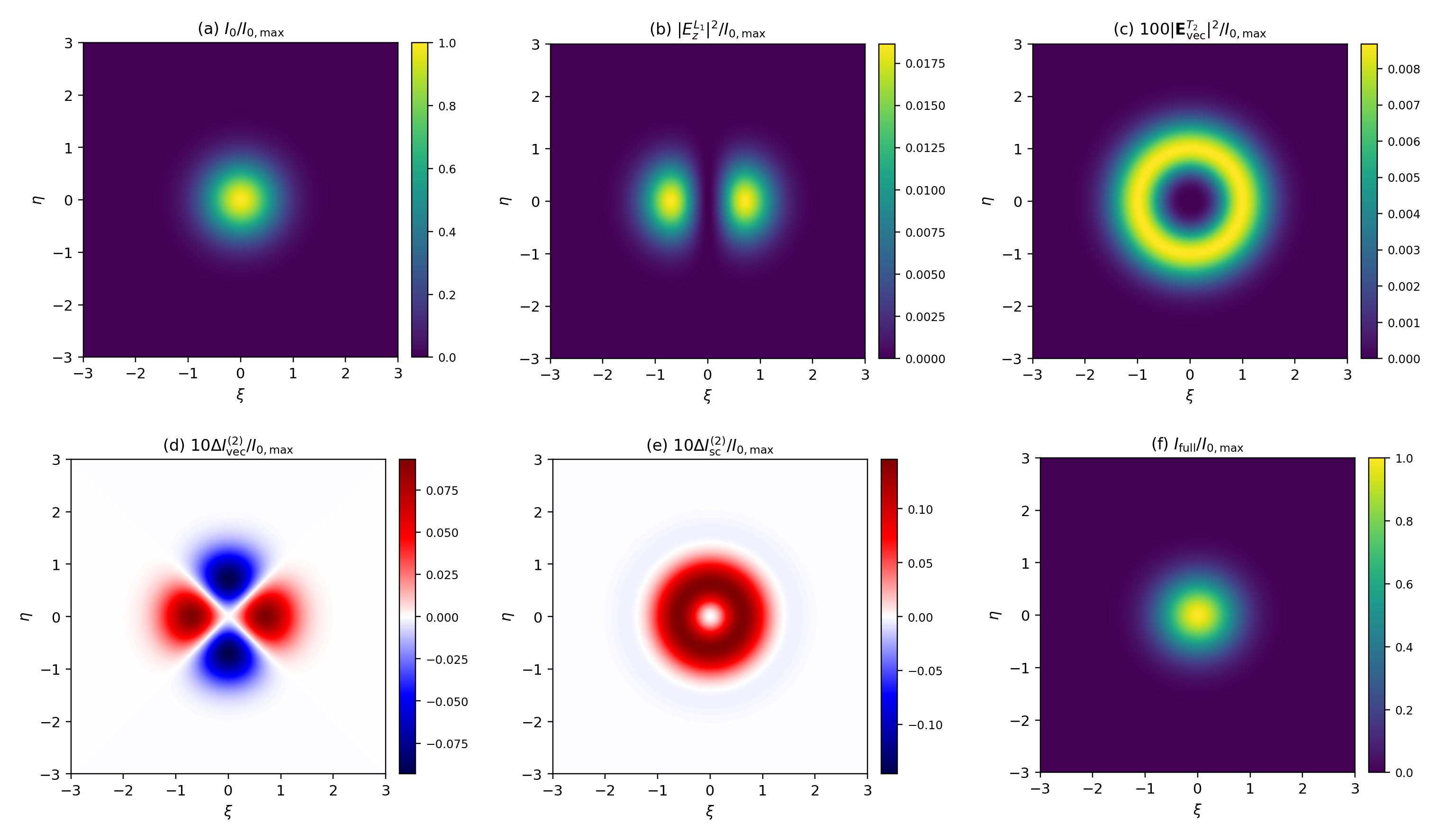}
    \caption{Intensity contributions for an \(x\)-polarized Gaussian beam at the waist, calculated from the scalar- and vectorial-corrected field.  The beam waist is chosen as \(w_0=\lambda\), corresponding to \(f=1/(2\pi)\). (a) Zeroth-order paraxial intensity \(I_0/I_{0,\max}\). (b) Longitudinal-field contribution \(|E_z^{L_1}|^2/I_{0,\max}\).  (c) Standalone second-order vectorial transverse contribution \(100|\mathbf{E}_{\rm vec}^{T_2}|^2/I_{0,\max}\).  (d) Interference between the paraxial field and the vectorial transverse correction, \(10\Delta I_{\rm vec}^{(2)}/I_{0,\max}\).  (e) Interference between the paraxial field and the scalar envelope correction, \(10\Delta I_{\rm sc}^{(2)}/I_{0,\max}\).  (f) Total intensity \(I_{\rm full}/I_{0,\max}\) obtained from the corrected field.  The signed interference terms in (d,e) are plotted on a red–blue scale centred at zero.
}
    \label{fig:figure1}
\end{figure}

For the particular choice $\psi_2=\psi_2^{(1)}$, one can recover the results presented in \cite{Barton_fifth_1989}. To illustrate the physical content of the different correction terms, Fig.~\ref{fig:figure1} shows the intensity contributions for an \(x\)-polarized Gaussian beam at the waist. The leading paraxial intensity remains dominant, so the full corrected intensity differs only weakly from the circular Gaussian profile. The individual interference terms, however, clearly reveal the structure of the scalar and vectorial corrections: the scalar correction is radially symmetric, whereas the vectorial correction produces an anisotropic four-lobed contribution. 

\begin{figure}[t]
    \centering
    \includegraphics[width=0.8\linewidth]{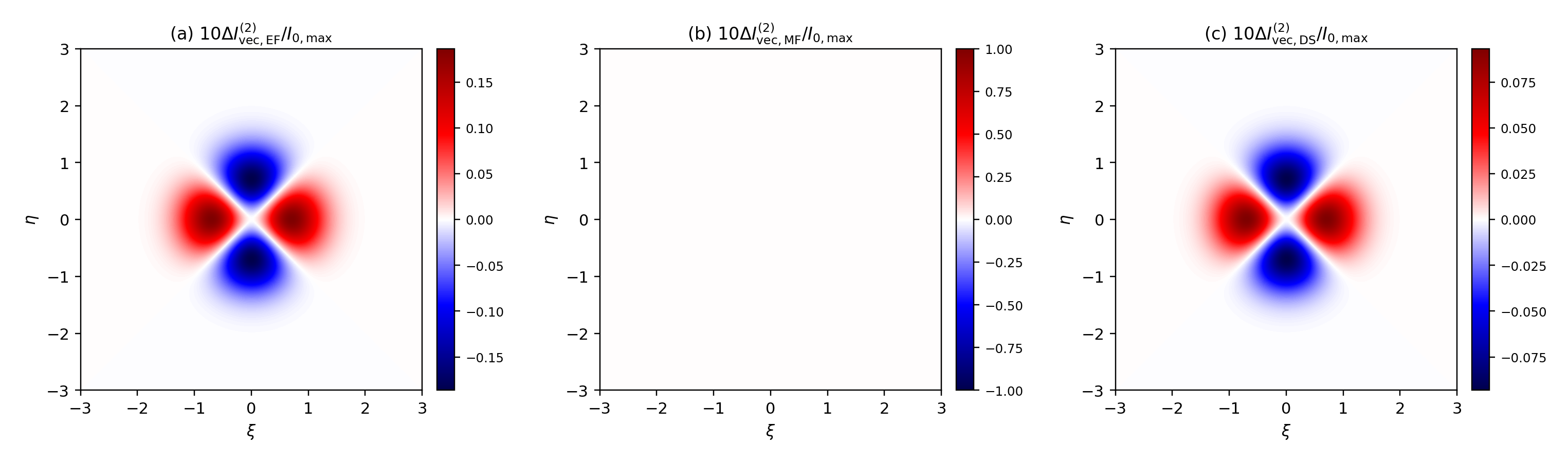}
   \caption{ Comparison of the second-order vectorial electric-intensity correction obtained from the electric-first, magnetic-first, and dual-symmetrized constructions for an \(x\)-polarized Gaussian beam at the waist.  The plotted quantity is \(10\Delta I_{\rm vec}^{(2)}/I_{0,\max}\), where \(\Delta I_{\rm vec}^{(2)}=2{\rm Re}[(\mathbf{E}^{T0})^*\cdot\mathbf{E}_{\rm vec}^{T2}]\).  (a) In the electric-first construction, the transverse vectorial \(T2\) correction is assigned entirely to the electric field, giving the full anisotropic interference pattern.  (b) In the magnetic-first construction, the corresponding vectorial \(T2\) correction is assigned to the magnetic field, so this electric-field interference contribution vanishes.  (c) In the dual-symmetrized construction, the correction is shared equally between the electric and magnetic fields, giving half the electric-first contribution.  This comparison illustrates that the ambiguity between one-sided iterative constructions does not affect the leading paraxial or longitudinal terms, but appears in the assignment of the second-order transverse vectorial correction.
 }
   \label{fig:figure2}
\end{figure}

Figure~\ref{fig:figure2} isolates the construction-dependent part of the theory. The leading paraxial field and the first longitudinal correction are common to the electric-first, magnetic-first and dual-symmetrized constructions. The difference appears in the second-order transverse vectorial correction. The electric-first construction assigns this correction to the electric field, the magnetic-first construction assigns the corresponding correction to the magnetic field, and the dual-symmetrized construction shares it equally between them.

\subsection{Circularly polarized LG beam}
The leading paraxial envelope of an LG beam with $p=0$ is expressed as:
\begin{equation}
\psi_{0}
=(\sqrt{2}\rho)^{|\ell|}
(-iQ)^{|\ell|+1}
e^{i\rho^{2}Q}
e^{i\ell\phi},
\end{equation}

The solution of $\psi_2$ satisfying $\psi_2(\rho,0)=0$ is given by \cite{Vovk_Analytical_2017}:
\begin{equation}
\psi_{2}^{(0)}
=\left[2i\zeta (|\ell|+1)(|\ell|+2)Q^{2}
-4\zeta (|\ell|+2)\rho^{2}Q^{3} -2i\zeta\rho^{4}Q^{4}\right]\psi_{0}.
\end{equation}

On the other hand, motivated by Davis’ asymptotic spherical-wave construction for the fundamental Gaussian beam, we seek an analogous particular solution for the $p=0$ LG mode:
\begin{equation}
\psi_{2}^{(1)}
=-i\left[(|\ell|+1)(|\ell|+2)Q+\rho^{4}Q^{3} \right]\psi_{0}.
\end{equation}

Both solutions are reduced to the expressions in the earlier subsection for $\ell=0$.

The corrected electric and magnetic fields are:
\begin{align}
\mathbf{E}_{\mathrm{corr}}
=
e^{i\zeta/f^{2}}
\Bigg\{&
\left[
\psi_{0}
+
f^{2}\left(\psi_{2}+\left[2i(|\ell|+1)Q-2\rho^{2}Q^{2}\right]\psi_{0}\right)\right]\mathbf{e}_{\sigma}\nonumber\\[1mm]
&+\frac{f^{2}}{4}\psi_{0}e^{2i\sigma\phi}\left[\frac{2(|\ell|-1)(|\ell|-\sigma\ell)}{\rho^{2}}+4i(|\ell|-\sigma\ell)Q-4\rho^{2}Q^{2}\right]\mathbf{e}_{-\sigma}\nonumber\\[1mm]
&+\frac{if}{\sqrt{2}}e^{i\sigma\phi}\left[\frac{|\ell|-\sigma\ell}{\rho}+2i\rho Q\right]\psi_{0}\,\hat{\mathbf{z}}\nonumber\\[1mm]
&+\frac{if^{3}}{\sqrt{2}}e^{i\sigma\phi}\left(\frac{\partial}{\partial\rho}
-\frac{\sigma\ell}{\rho}\right)\left\{\psi_{2}+\left[i(|\ell|+1)Q-\rho^{2}Q^{2}\right]\psi_{0}\right\}\hat{\mathbf{z}}\Bigg\}.
\end{align}

and,

\begin{align}
\mathbf{B}_{\mathrm{corr}}
=-\frac{i\sigma}{c}e^{i\zeta/f^{2}}\Bigg\{&\left[\psi_{0}
+f^{2}\left(\psi_{2}+\left[2i(|\ell|+1)Q-2\rho^{2}Q^{2}\right]\psi_{0}\right)\right]\mathbf{e}_{\sigma}\nonumber\\[1mm]
&+\frac{f^{2}}{4}\psi_{0}e^{2i\sigma\phi}\left[\frac{2(|\ell|-1)(|\ell|-\sigma\ell)}{\rho^{2}}+4i(|\ell|-\sigma\ell)Q-4\rho^{2}Q^{2}\right]\mathbf{e}_{-\sigma}\nonumber\\[1mm]
&+\frac{if}{\sqrt{2}}e^{i\sigma\phi}\left[\frac{|\ell|-\sigma\ell}{\rho}
+2i\rho Q\right]\psi_{0}\,\hat{\mathbf{z}}\nonumber\\[1mm]
&+\frac{if^{3}}{\sqrt{2}}e^{i\sigma\phi}\left(\frac{\partial}{\partial\rho}-\frac{\sigma\ell}{\rho}\right)\left\{\psi_{2}+\left[i(|\ell|+1)Q-\rho^{2}Q^{2}\right]\psi_{0}\right\}\hat{\mathbf{z}}\Bigg\}.
\end{align}

The present formulation brings together the slow longitudinal-envelope contribution, scalar-envelope corrections, and electric–magnetic symmetrization in a unified treatment of LG beams. These ingredients are not incorporated jointly in the specific formulations of Refs \cite{Alharbi_role_2025,Forbes_VortexLightNanoscale_2025,Vovk_Analytical_2017}.
In particular, for the choice $\psi_2=\psi_2^{(0)}$, the corresponding
unsymmetrized electric-first expressions can be mapped onto the
perturbative LG fields presented in
\cite{Vovk_Analytical_2017} through the common truncation order.

%These fields are dual symmetrized and represent more accurate and complete expressions compared to literature \cite{Alharbi_role_2025,Forbes_VortexLightNanoscale_2025,Vovk_Analytical_2017}. For example, the corresponding unsymmetrized electric-first expressions for the particular choice $\psi_2=\psi_2^{0}$ can be mapped onto the perturbative LG fields presented in \cite{Vovk_Analytical_2017}.

\section{Conclusion}

Within the present formulation, the higher-order nonparaxial corrections are naturally separated into scalar and vectorial contributions.
The scalar term modifies the main beam envelope, whereas the vectorial Maxwell correction introduces additional transverse and longitudinal structures. 
By consistently retaining the slow longitudinal derivative of the paraxial envelope and implementing electric–magnetic symmetrization, the proposed scheme treats these corrections symmetrically and avoids assigning the full vectorial correction to only one field.
The presented formulation provides a transparent and readily extensible route for constructing symmetrized nonparaxial fields from a general paraxial envelope and input polarization. In particular, we have applied the formulation to ($p=0$) Laguerre–Gaussian beams, for which higher-order corrections have attracted considerable recent interest. The resulting framework therefore provides a consistent theoretical basis for assessing their influence on polarization structure, intensity distributions, spin–orbit effects, and light–matter interactions.

Although perturbative schemes are powerful, they must be interpreted carefully; they should be understood as asymptotic rather than truly convergent. Gouesbet and co-workers showed, for example, that the Davis-Barton scheme is divergent at sufficiently high order, despite being highly useful at relatively low orders \cite{gouesbet_Divergence_2020}. 
The perturbative expansion should be also used with caution as
the focusing strength increases and the expansion parameter $f$ is no longer small.
Establishing the range of $f$ over which the present formulation remains accurate, through comparison with fully nonparaxial treatment, is left for future work.

\bibliography{References2026}

\end{document}